\documentstyle[preprint,eqsecnum,aps]{revtex}

\begin{document}
\draft
\tightenlines
\preprint{
\rightline{\vbox{\hbox{\rightline{MSUCL-910}}
\hbox{\rightline{McGill/93-41}}}}
         }
\title{Properties of the phi meson at finite temperature}
\author{Kevin L. Haglin\cite{kevin}}
\address{
National Superconducting Cyclotron Laboratory, Michigan
State University\\
East Lansing, Michigan 48824--1321
}
\author{Charles Gale\cite{charles}}
\address{
Physics Department, McGill University, 3600 University Street\\
Montr\'eal, QC H3A 2T8,  Canada
}
\date{\today}
\maketitle
\begin{abstract}
We calculate the $\phi$-meson propagator at finite temperature at the one--loop
order.
The real and imaginary parts are studied separately in full
kinematic ranges.  From this activity we
investigate how temperature affects such things as
decay widths and dispersion relations.
{}From here we estimate the thermal rate of lepton pair radiation in a hadron
gas proceeding through
$K^{+}K^{-}\to \phi \to \ell^{+}\ell^{-}$ and
$\pi\rho\to \phi \to \ell^{+}\ell^{-}$.
We find several interesting things.  From the dispersion
relations we learn the effective mass calculated this way
increases with temperature as does the partial width, but only slightly.
At $T=200$ MeV, the mass increases by $\sim$ 4 MeV and the partial
width increases by 34\%.  Polarizations are indistinguishable for practical
purposes.
\end{abstract}

\narrowtext

\section{Introduction}
\label{sec:intro}

Attention is focused nowadays
on problems relating to the physics of ultra-relativistic heavy-ion
collisions where the primary purpose is to establish whether or
not one can make a quark-gluon plasma (QGP).  Regardless of the outcome
of this query, the hadronic phase is present and it behooves us to
continue thought and calculation regarding hadronic mechanisms
in this extremely hot environment.
It is clear that hadronic properties are modified at finite
temperature.  But to what extent such modifications are noticeable
and for which hadrons this becomes important is not so easy to say.  By
studying these modifications we may be able to better understand what is
happening as temperatures approach that which is expected to give a phase
transition to the chirally symmetric QGP phase.   There have been several
studies of meson masses at finite temperature.  Some
employ lattice QCD (quantum chromodynamic) methods \cite{cdet85,sgot87},
and others use QCD sum rules \cite{mdey90,rfur90}.  Still others
have used effective Lagrangian methods \cite{cgal91,cson93}.  Our choice is
the latter which uses finite-temperature field theory.

There are more than a half dozen hadrons with masses
less then $\sim$ 1 GeV.  At the upper end of that range sits
the $\phi$-meson.  It is a well known two-kaon resonance that
is less abundant in these heavy-ion collisions than $\rho$\,s or
their decay products, the pions.  On the other hand, the invariant
mass distribution of lepton pairs is expected to show a noticeable
structure near the $\phi$'s peak \cite{cgal93} so it is important
at some level.
Its finite temperature behavior
including two-kaon annihilation into
lepton pairs has been studied \cite{cgal91}.
Deviation from vacuum behavior was concluded to be quite small.
Yet, this was in some sense not surprising since the kaons are
3--4 times more massive than temperatures of interest.
Again, the $\phi$-meson decays into two kaons with
highest rate but only somewhat less often decays into a $\pi\rho$ combination.
The combined
mass of the $\pi\rho$ system is smaller than the two-kaon threshold but
not by enough to argue that the rate is sizeable merely due to phase space.
The pion in the $\pi\rho$ decay of the $\phi$ might be more strongly
affected by temperature than the kaons since it much less massive.   Of
course, the opposite argument applies to the $\rho$ being more massive than
the kaons, so {\em a priori} it is not clear
what will happen to the whole system.  The questions we attempt to
answer are the following.  How does the $\pi\rho$ decay possibility
affect the $\phi$'s zero and finite temperature behavior?  And, how does
it affect the dispersion relation, i.e. does the effective mass remain the
same, go up or down?  The interesting role played by $\phi$ mesons and they
lepton pair decay channel has previously been pointed in the context of high
energy heavy ion collisions \cite{shurlis}.

Our paper is organized in the following way.  In Sec.\ \ref{sec:selfenergy}
we briefly discuss the one-loop kaon diagrams' contributions to the $\phi$
self-energy.  They have been computed and
appear in the literature \cite{cgal91}.   We then evaluate
the $\pi\rho$ loop's contribution and add it to the kaon diagrams.
With all these, we have an improved $\phi$-meson self-energy.
It is a simple task to get from their to the full propagator which we
present in Sec.~{\ref{sec:propagator}.  Many properties of the resonance
are contained within the propagator.  Among them, we present effective
widths and masses and dispersion relations.  We finish with a short
discussion on the thermal dilepton emission rates
which are evaluated from knowledge of the finite temperature behavior
of the imaginary part of the propagator.
Finally, in Sec.\ \ref{sec:conclusions}
we summarize and conclude that temperature effects are quite modest in this
improved $\phi$-meson propagator.

\section{$\phi$-meson self-energy to one-loop}
\label{sec:selfenergy}

Vector mesons interact in a renormalizable fashion with a conserved
current.  The full Lagrangian for $K$-$\phi$
dynamics is
\begin{equation}
{\cal L} = {1\over 2}|D_{\mu}\bbox{K}|^{2}-{1\over 2}m_{K}^{2}
|\bbox{K}|^{2}
-{1\over 4}\phi_{\mu\nu}\phi^{\mu\nu} + {1\over 2}m_{\phi}^{2}\phi_{\mu}
\phi^{\mu},
\label{eq:lagrangian1}
\end{equation}
where $\bbox{K}$ is the complex charged kaon field, $\phi_{\mu\nu}=
\partial_{\mu}\phi_{\nu}-\partial_{\nu}\phi_{\mu}$ is the $\phi$ field
strength tensor, and $D_{\mu}=\partial_{\mu}-ig_{\phi KK}\phi_{\mu}$ is the
covariant derivative.  This gives rise to two diagrams which contribute
to the self-energy at the one-loop level shown in
Figs.~\ref{fig:diagrams}$a$ and \ref{fig:diagrams}$b$.
We go further by including
a $\phi$-$\rho$-$\pi$ interaction given by the Lagrangian \cite{umei87}
\begin{equation}
{\cal L}_{\rm int} = g_{\phi\rho\pi}\epsilon_{\mu\nu\alpha\beta}
\partial^{\mu}\phi^{\nu} \partial^{\alpha}\bbox{\rho}^{\beta}
\cdot \bbox{\pi},
\label{eq:lagrangian2}
\end{equation}
where the complex charged rho and pion fields appears as $\bbox{\rho}^{\beta}$
and $\bbox{\pi}$.
This term gives the diagram shown in Fig.~\ref{fig:diagrams}$c$.
The kaon bubble and tadpole diagrams have been evaluated (by one of us along
with J. Kapusta) and appear in the literature\cite{cgal91}.   So we discuss
only the $\pi\rho$ loop diagram.  Its contribution to
the self-energy in Euclidean space is
\begin{equation}
\Pi^{\mu\nu}(k) = 3g_{\phi\rho\pi}^{2}\;T \sum\limits_{p_{4}} \int
{d^{3}p \over (2\pi)^{3}}
{{\cal I}^{\mu\nu}
\over (p^{2}+m_{\pi}^{2})\left\lbrack(p+k)^{2}+m_{\rho}^{2}\right\rbrack}
\label{eq:selfenergy}
\end{equation}
where
\begin{equation}
{\cal I}^{\mu\nu} = {
\lbrack (k\cdot p)^{2} - p^{2}k^{2} \rbrack \delta^{\mu\nu}+
p^{2} k^{\mu}k^{\nu}+ k^{2}p^{\mu}p^{\nu}-(k\cdot p)
\lbrack p^{\mu}k^{\nu} + k^{\mu}p^{\nu} \rbrack }.
\label{eq:numerator}
\end{equation}
The factor of 3 accounts for a sum over loop isospin.
In the imaginary-time formalism, the indices run not from 0 to 3 but 1
to 4.  The fourth component of the four-vectors
are Matsubara frequencies, $p_{4}$ and $k_{4}$ = $2\pi T \times$ integer.

As with all imaginary-time calculations, the zero temperature piece
naturally separates from the $T$ dependent piece.  We
have evaluated the expression in Eqs.~(\ref{eq:selfenergy}) and
(\ref{eq:numerator}) at $T=0$ using
dimensional regularization \cite{pram}.  The rest, which depends
on temperature, is obtained by
converting the discrete sum over frequencies to a contour integral.  We
find it convenient to present the zero temperature results separately.
For this $\pi$-$\rho$ loop we obtain
\begin{eqnarray}
\Pi^{\mu\nu}_{\rm vac}(k) &=& {g_{\phi \rho\pi}^{2} \over 16\pi^{2}}
\left(k^{\mu}k^{\nu}-k^{2}\delta^{\mu\nu}\right)
\left\lbrace {1\over 4 k^{2}}
\left( (m_{\rho}^{2}-m_{\pi}^{2}+k^{2})^{2}
+4m_{\pi}^{2}k^{2} \right)\sqrt{\cal R}
                                         \right.
\nonumber\\
& &\quad\times
\ln { \left(1+{m_{\rho}^{2}-m_{\pi}^{2}\over k^{2}}+\sqrt{\cal R}\right)
      \left(1-{m_{\rho}^{2}-m_{\pi}^{2}\over k^{2}}+\sqrt{\cal R}\right)
\over
      \left(1+{m_{\rho}^{2}-m_{\pi}^{2}\over k^{2}}-\sqrt{\cal R}\right)
      \left(1-{m_{\rho}^{2}-m_{\pi}^{2}\over k^{2}}-\sqrt{\cal R}\right)
    } \nonumber\\
& &\quad  -{1\over 2k^{4}}(m_{\rho}^{2}-m_{\pi}^{2}+k^{2})\left[
\left({m_{\rho}^{2}-m_{\pi}^{2}+k^{2}}\right)^{2}+6m_{\pi}^{2}k^{2}
\right]
\ln \left( {m_{\rho}\over m_{\pi}}\right) \nonumber\\
& &\quad \left. -{5\over 6}k^{2} - {(m_{\rho}^{2}-m_{\pi}^{2})^{2}\over
2k^{2}} -{2}\left(m_{\rho}^{2}+m_{\pi}^{2}\right)
+C\left(3m_{\rho}^{2}+3m_{\pi}^{2}+{k^{2}} \right) \right\rbrace
\label{eq:pimunuvac}
\end{eqnarray}
at $T=0$, with
\begin{equation}
{\cal R} = \left(1 + {m_{\rho}^{2}-m_{\pi}^{2} \over k^{2}}
\right)^{2} + {4m_{\pi}^{2} \over k^{2}},
\end{equation}
and where $C$ is a renormalization constant
to be determined later.  Then at $T>0$
\begin{eqnarray}
\Pi^{44}_{\rm mat}(k) &=& 3\left({g_{\phi\rho\pi}\over 2\pi}\right)^{2}
\int\limits_{0}^{\infty}dp \, p^{2} \left\lbrace {1 \over \omega_{\pi}}
\left[{-(k_{4}^{2}+\bbox{k}^{2}+m_{\rho}^{2}-m_{\pi}^{2})\over 2}
                            \right. \right.
\nonumber\\
& & + {4\omega_{\pi}^{2}k_{4}^{2}+4p^{2}\bbox{k}^{2}-(k_{4}^{2}+
\bbox{k}^{2}+m_{\rho}^{2}-m_{\pi}^{2})^{2} \over 16p|\bbox{k}| }\ln a_{\pi}
\nonumber\\
& & + \left.
{i\omega_{\pi}k_{4}(k_{4}^{2}+\bbox{k}^{2}+m_{\rho}^{2}-m_{\pi}^{2})
\over 4p|\bbox{k}|  }
\ln b_{\pi} \right]
{1 \over e^{\beta\omega_{\pi}}-1} \nonumber\\
& &\quad\quad\quad\quad\quad\quad\quad\quad + {1 \over \omega_{\rho}}
\left[{-(k_{4}^{2}+\bbox{k}^{2}+m_{\pi}^{2}-m_{\rho}^{2})\over 2}
                           \right.
\nonumber\\
& & + {4\omega_{\rho}^{2}k_{4}^{2}+4p^{2}\bbox{k}^{2}-(k_{4}^{2}+
\bbox{k}^{2}+m_{\pi}^{2}-m_{\rho}^{2})^{2} \over 16p|\bbox{k}| }\ln a_{\rho}
\nonumber\\
& & + \left. \left.
{i\omega_{\rho}k_{4}(k_{4}^{2}+\bbox{k}^{2}+m_{\pi}^{2}-m_{\rho}^{2})
\over 4p|\bbox{k}|  }
\ln b_{\rho} \right]
{1 \over e^{\beta\omega_{\rho}}-1} \right\rbrace,
\label{eq:pi44}
\end{eqnarray}
\begin{equation}
\Pi^{4i}_{\rm mat}(k) = -{k_{4}k^{i}\over \bbox{k}^{2}} \Pi^{44}_{\rm mat}(k),
\end{equation}
and
\begin{equation}
\Pi^{ij}_{\rm mat}(k) = A\delta^{ij} + {k^{i}k^{j}\over \bbox{k}^{2}}B
\end{equation}
where
\begin{eqnarray}
A &=& 3\left({g_{\phi\rho\pi}\over 2\pi}\right)^{2}
\int\limits_{0}^{\infty}dp \, p^{2} \left\lbrace {1 \over \omega_{\pi}}
\left[{(k_{4}^{2}-\bbox{k}^{2})(k_{4}^{2}+\bbox{k}^{2}+m_{\rho}^{2}
-m_{\pi}^{2})\over 4\bbox{k}^{2}}
            \right. \right.
\nonumber\\
& & + \left({(k_{4}^{2}-\bbox{k}^{2})
\left((k_{4}^{2}+\bbox{k}^{2}+m_{\rho}^{2}-m_{\pi}^{2})^{2}
-4\omega_{\pi}^{2}k_{4}^{2}\right) \over 32p|\bbox{k}|^{3}} -
{\omega_{\pi}^{2}k_{4}^{2}\over 2p|\bbox{k}|} \right. \nonumber\\
& & \left. -{p^{2}(\bbox{k}^{2}-k_{4}^{2})+2m_{\pi}^{2}\bbox{k}^{2}\over
8p|\bbox{k}|}
\right)\ln a_{\pi}
\nonumber\\
& & - \left.
{i\omega_{\pi}k_{4}(k_{4}^{2}+\bbox{k}^{2})(k_{4}^{2}+\bbox{k}^{2}
+m_{\rho}^{2}-m_{\pi}^{2})\over 8p|\bbox{k}|^{3}  }
\ln b_{\pi} \right]
{1 \over e^{\beta\omega_{\pi}}-1} \nonumber\\
& &\quad\quad\quad\quad\quad\quad\quad\quad + {1 \over \omega_{\rho}}
\left[{(k_{4}^{2}-\bbox{k}^{2})(k_{4}^{2}+\bbox{k}^{2}+m_{\pi}^{2}-m_{\rho}^{2})
\over 4\bbox{k}^{2}}
          \right.
\nonumber\\
& & + \left({(k_{4}^{2}-\bbox{k}^{2})
\left((k_{4}^{2}+\bbox{k}^{2}+m_{\pi}^{2}-m_{\rho}^{2})^{2}
-4\omega_{\rho}^{2}k_{4}^{2}\right) \over 32p|\bbox{k}|^{3}} -
{\omega_{\rho}^{2}k_{4}^{2}\over 2p|\bbox{k}|} \right. \nonumber\\
& & \left. -{p^{2}(\bbox{k}^{2}-k_{4}^{2})+2m_{\rho}^{2}\bbox{k}^{2}\over
8p|\bbox{k}|}
\right)\ln a_{\rho}
\nonumber\\
& & - \left. \left.
{i\omega_{\rho}k_{4}(k_{4}^{2}+\bbox{k}^{2})(k_{4}^{2}+\bbox{k}^{2}
+m_{\pi}^{2}-m_{\rho}^{2})\over 8p|\bbox{k}|^{3}  }
\ln b_{\rho} \right]
{1 \over e^{\beta\omega_{\rho}}-1}  \right\rbrace,
\label{eq:a}
\end{eqnarray}
\begin{eqnarray}
B &=& 3\left({g_{\phi\rho\pi}\over 2\pi}\right)^{2}
\int\limits_{0}^{\infty}dp \, p^{2} \left\lbrace {1 \over \omega_{\pi}}
\left[{-(3k_{4}^{2}-\bbox{k}^{2})(k_{4}^{2}+\bbox{k}^{2}
+m_{\rho}^{2}-m_{\pi}^{2})\over 4\bbox{k}^{2}}
         \right. \right.
\nonumber\\
& & - \left({(3k_{4}^{2}-\bbox{k}^{2})
\left((k_{4}^{2}+\bbox{k}^{2}+m_{\rho}^{2}-m_{\pi}^{2})^{2}
-4\omega_{\pi}^{2}k_{4}^{2}\right) \over 32p|\bbox{k}|^{3}} -
{\omega_{\pi}^{2}k_{4}^{2}\over 2p|\bbox{k}|} \right.\nonumber\\
& & \left. -{m_{\pi}^{2}|\bbox{k}|\over 4p} -{p(k_{4}^{2}+\bbox{k}^{2})\over
8|\bbox{k}|}
\right)\ln a_{\pi}
\nonumber\\
& & \left.
+{i\omega_{\pi}k_{4}(3k_{4}^{2}+\bbox{k}^{2})(k_{4}^{2}+\bbox{k}^{2}
+m_{\rho}^{2}-m_{\pi}^{2})\over 8p|\bbox{k}|^{3}  }
\ln b_{\pi} \right]
{1 \over e^{\beta\omega_{\pi}}-1} \nonumber\\
& &\quad\quad\quad\quad\quad\quad\quad\quad + {1 \over \omega_{\rho}}
\left[{-(3k_{4}^{2}-\bbox{k}^{2})(k_{4}^{2}+\bbox{k}^{2}
+m_{\pi}^{2}-m_{\rho}^{2})\over 4\bbox{k}^{2}}
         \right.
\nonumber\\
& & - \left({(3k_{4}^{2}+\bbox{k}^{2})
\left((k_{4}^{2}+\bbox{k}^{2}+m_{\pi}^{2}-m_{\rho}^{2})^{2}
-4\omega_{\rho}^{2}k_{4}^{2}\right) \over 32p|\bbox{k}|^{3}} \right.\nonumber\\
& & \left. -{\omega_{\rho}^{2}k_{4}^{2}\over 2p|\bbox{k}|} -
{m_{\rho}^{2}|\bbox{k}|\over 4p}
-{p(k_{4}^{2}+\bbox{k}^{2})\over 8|\bbox{k}|}
\right)\ln a_{\rho}
\nonumber\\
& & + \left. \left.
{i\omega_{\pi}k_{4}(3k_{4}^{2}+\bbox{k}^{2})(k_{4}^{2}+\bbox{k}^{2}+
m_{\pi}^{2}-m_{\rho}^{2})\over 8p|\bbox{k}|^{3}  }
\ln b_{\rho} \right]
{1 \over e^{\beta\omega_{\rho}}-1}  \right\rbrace,
\label{eq:b}
\end{eqnarray}
and finally,
\begin{eqnarray}
a_{\pi} &=& {(k_{4}^{2}+\bbox{k}^{2}+m_{\rho}^{2}-m_{\pi}^{2}
-2p|\bbox{k}|)^{2}+
4\omega_{\pi}^{2}k_{4}^{2} \over (k_{4}^{2}+\bbox{k}^{2}+m_{\rho}^{2}-
m_{\pi}^{2}+2p|\bbox{k}|)^{2}+4\omega_{\pi}^{2}k_{4}^{2}}, \nonumber\\
a_{\rho} &=& {(k_{4}^{2}+\bbox{k}^{2}+m_{\pi}^{2}-m_{\rho}^{2}
-2p|\bbox{k}|)^{2}+
4\omega_{\rho}^{2}k_{4}^{2} \over (k_{4}^{2}+\bbox{k}^{2}+m_{\pi}^{2}-
m_{\rho}^{2}+2p|\bbox{k}|)^{2}+4\omega_{\rho}^{2}k_{4}^{2}}, \nonumber\\
b_{\pi} &=& {(k_{4}^{2}+\bbox{k}^{2}+m_{\rho}^{2}-m_{\pi}^{2})^{2}
-4(p|\bbox{k}|+
i\omega_{\pi}k_{4})^{2} \over (k_{4}^{2}+\bbox{k}^{2}+m_{\rho}^{2}-
m_{\pi}^{2})^{2}-4(p|\bbox{k}|-i\omega_{\pi}k_{4})^{2}}, \nonumber\\
b_{\rho} &=& {(k_{4}^{2}+\bbox{k}^{2}+m_{\pi}^{2}-m_{\rho}^{2})^{2}
-4(p|\bbox{k}|+i\omega_{\rho}k_{4})^{2} \over (k_{4}^{2}+\bbox{k}^{2}+
m_{\pi}^{2}-
m_{\rho}^{2})^{2}-4(p|\bbox{k}|-i\omega_{\rho}k_{4})^{2}},
\label{eq:aandb}
\end{eqnarray}
where $\omega_{\pi}=\sqrt{p^{2}+m_{\pi}^{2}}$ and
$\omega_{\rho}=\sqrt{p^{2}+m_{\rho}^{2}}$.
We have evaluated these components separately and then checked
for transversality as required by current conservation.

\section{Full propagator}
\label{sec:propagator}

We next let time become real which changes the space-time
from Euclidean to Minkowskian.  The metric is now
$+\, -\, -\, - $ for $\mu=0,1,2,3$.  We can decompose the self-energy
into scalar functions times longitudinal and transverse projection
tensors defined to be
\begin{eqnarray}
P_{\rm T}^{00}&=& P_{\rm T}^{0i} = P_{\rm T}^{i0}=0, \nonumber\\
P_{\rm T}^{ij} &=& \delta^{ij}-k^{i}k^{j}/\bbox{k}^{2}, \nonumber\\
P_{\rm L}^{\mu\nu}&=&k^{\mu}k^{\nu}/k^{2}-g^{\mu\nu}-P_{\rm T}^{\mu\nu}.
\end{eqnarray}
The decomposition reads
\begin{equation}
\Pi^{\mu\nu}(k) = F(k)P_{\rm L}^{\mu\nu}+G(k)P_{\rm T}^{\mu\nu}.
\end{equation}
Utilizing this, we have
\begin{equation}
F = -{k^{2}\over \bbox{k}^{2}}\Pi^{44}, \quad\quad G=A,
\end{equation}
where $\Pi^{44}$ and $A$ are given by Eqs.~(\ref{eq:pi44}) and (\ref{eq:a})
with $ik_{4}=k_{0}$.  Inner-products are again Minkowskian, e.g.
$k^{2}=k_{0}^{2}-\bbox{k}^{2}$.
The all-important connection with the full and bare propagator comes next,
it is
\begin{equation}
\Pi^{\mu\nu} = \left({\cal D}^{\mu\nu}\right)^{-1}-
\left({\cal D}^{\mu\nu}_{0}\right)^{-1}.
\end{equation}
Therefore, the full $\phi$-meson propagator can be written as
\begin{equation}
{\cal D}^{\mu\nu} = -{ P_{\rm L}^{\mu\nu} \over k^{2}-m_{\phi}^{2}-F}
-{ P_{\rm T}^{\mu\nu} \over k^{2}-m_{\phi}^{2}-G} -
{k^{\mu}k^{\nu}\over k^{2}m_{\phi}^{2}}.
\end{equation}
The first two pieces represent the longitudinal and transverse collective
excitations of the resonance at finite temperature and in general, are
different.
To learn just how different they are for $T>0$ is now a
numerical task which will be discussed later.

In the vacuum there is no preferred rest frame and the scalar functions
$F$ and $G$ become equal.  After dimensional regularization and
renormalization we arrive at finite results
\begin{eqnarray}
F_{\rm  vac} &=& G_{\rm vac}=
{g_{\phi \rho\pi}^{2} \over 16\pi^{2}}M^{2}
\left\lbrace {-1\over 4 M^{2}}
\left( (m_{\rho}^{2}-m_{\pi}^{2}-M^{2})^{2} - 4m_{\pi}^{2}M^{2}
\right)\sqrt{\cal R}
               \right.
\nonumber\\
& &\quad\times \left[
\ln \left\vert
    { \left(1-{m_{\rho}^{2}-m_{\pi}^{2}\over M^{2}}+\sqrt{\cal R}\right)
      \left(1+{m_{\rho}^{2}-m_{\pi}^{2}\over M^{2}}+\sqrt{\cal R}\right)
\over
      \left(1-{m_{\rho}^{2}-m_{\pi}^{2}\over M^{2}}-\sqrt{\cal R}\right)
      \left(1+{m_{\rho}^{2}-m_{\pi}^{2}\over M^{2}}-\sqrt{\cal R}\right)
    } \right\vert
+i\pi\Theta\left(M^{2}-(m_{\pi}+m_{\rho})^{2}\right)
\right] \nonumber\\
& &\quad  -{1\over 2M^{4}}(m_{\rho}^{2}-m_{\pi}^{2}-M^{2})\left[
\left({m_{\rho}^{2}-m_{\pi}^{2}-M^{2}}\right)^{2}
-6m_{\pi}^{2}M^{2}
\right]
\ln \left( {m_{\rho}\over m_{\pi}}\right) \nonumber\\
& &\quad \left. +{5\over 6}M^{2} + {(m_{\rho}^{2}-m_{\pi}^{2})^{2}\over
2M^{2}} -{2}\left(m_{\rho}^{2}+m_{\pi}^{2}\right)
+C \left(3m_{\rho}^{2}+3m_{\pi}^{2}-{M^{2}} \right) \right\rbrace
\label{eq:fvac}
\end{eqnarray}
where now the radicand is
\begin{equation}
{\cal R} = \left(1 - {m_{\rho}^{2}-m_{\pi}^{2} \over M^{2}}
\right)^{2} - {4m_{\pi}^{2} \over M^{2}},
\end{equation}
and the invariant mass of the resonance is $M = \sqrt{k^{2}}$
while its energy is $E=\sqrt{M^{2}+\bbox{k}^{2}}$ in the rest frame
of the hadron gas.
We determine the renormalization constant by imposing the
physical-mass condition, namely,
Re\,$F_{\rm vac}(k^{2}=m_{\phi}^{2})=0$.
It determines the constant to be
\begin{eqnarray}
C &=& {1\over \left(3m_{\rho}^{2}+3m_{\pi}^{2}-{m_{\phi}^{2}} \right)}
 \left\lbrace   {1\over 4m_{\phi}^{2}}\left(
(m_{\rho}^{2}-m_{\pi}^{2}-m_{\phi}^{2})^{2}-4m_{\pi}^{2}m_{\phi}^{2}
\right)\sqrt{\cal R^{\prime}} \right.
\nonumber\\
& &\quad \times \ln \left\vert { \left(1-{m_{\rho}^{2}-m_{\pi}^{2}\over
m_{\phi}^{2}}+\sqrt{\cal R^{\prime}}\right)
      \left(1+{m_{\rho}^{2}-m_{\pi}^{2}\over m_{\phi}^{2}}
+\sqrt{\cal R^{\prime}}
\right)
\over
      \left(1-{m_{\rho}^{2}-m_{\pi}^{2}\over m_{\phi}^{2}}
-\sqrt{\cal R^{\prime}}\right)
      \left(1+{m_{\rho}^{2}-m_{\pi}^{2}\over m_{\phi}^{2}}
-\sqrt{\cal R^{\prime}}
\right)
    } \right\vert \nonumber\\
& & +{1\over 2m_{\phi}^{4}}(m_{\rho}^{2}-m_{\pi}^{2}-m_{\phi}^{2})
\left[\left({m_{\rho}^{2}-m_{\pi}^{2}-m_{\phi}^{2}}\right)^{2}
-6m_{\pi}^{2}m_{\phi}^{2} \right]\ln\left({m_{\rho}\over
m_{\pi}}\right) \nonumber\\
& &\left. -{5 \over 6}m_{\phi}^{2} - {(m_{\rho}^{2}-m_{\pi}^{2})^{2}
\over 2m_{\phi}^{2}}+{2}\left( m_{\rho}^{2}+m_{\pi}^{2}\right)
\right\rbrace
\end{eqnarray}
which use yet a different radicand
\begin{equation}
{\cal R^{\prime}} = \left(1-{m_{\rho}^{2}-m_{\pi}^{2}\over m_{\phi}^{2}}
\right)^{2} - {4m_{\pi}^{2}\over m_{\phi}^{2}}.
\end{equation}
At finite temperature the functions $F$ and $G$ have---in addition to
the vacuum piece---a piece due to the matter $F=F_{\rm vac}+F_{\rm mat}$ and
$G=G_{\rm vac}+G_{\rm mat}$, where
\begin{eqnarray}
F_{\rm mat} &=& -3{g_{\phi\rho\pi}^{2}\over 4\pi^{2}}
{M^{2}\over \bbox{k}^{2}}
\int\limits_{0}^{\infty}dp \, p^{2} \left\lbrace {1 \over \omega_{\pi}}
\left[{(M^{2}-m_{\rho}^{2}+m_{\pi}^{2})\over 2}
                            \right. \right.
\nonumber\\
& & + {-4\omega_{\pi}^{2}E^{2}+4p^{2}\bbox{k}^{2}-(M^{2}-
m_{\rho}^{2}+m_{\pi}^{2})^{2} \over 16p|\bbox{k}| }[\ln |a_{\pi}|
+i\pi{\cal B}_{\pi}]
\nonumber\\
& & + \left.
{\omega_{\pi}E(-M^{2}+m_{\rho}^{2}-m_{\pi}^{2})
\over 4p|\bbox{k}|  }
[\ln |b_{\pi}|-i\pi{\cal B}_{\pi}] \right]
{1 \over e^{\beta\omega_{\pi}}-1} \nonumber\\
& &\quad\quad\quad\quad\quad\quad\quad\quad + {1 \over \omega_{\rho}}
\left[{(M^{2}-m_{\pi}^{2}+m_{\rho}^{2})\over 2}
                           \right.
\nonumber\\
& & + {-4\omega_{\rho}^{2}E^{2}+4p^{2}\bbox{k}^{2}-(M^{2}-
m_{\pi}^{2}+m_{\rho}^{2})^{2} \over 16p|\bbox{k}| }[\ln |a_{\rho}|
+i\pi{\cal B}_{\rho}]
\nonumber\\
& & + \left. \left.
{\omega_{\rho}E(-M^{2}+m_{\pi}^{2}-m_{\rho}^{2})
\over 4p|\bbox{k}|  }
[\ln |b_{\rho}| -i\pi{\cal B}_{\rho}]
 \right]
{1 \over e^{\beta\omega_{\rho}}-1} \right\rbrace,
\label{eq:fmat}
\end{eqnarray}
\begin{eqnarray}
G_{\rm mat} &=& -3{g_{\phi\rho\pi}^{2}\over 4\pi^{2}}
\int\limits_{0}^{\infty}dp \, p^{2} \left\lbrace {1 \over \omega_{\pi}}
\left[{-(E^{2}+\bbox{k}^{2})(M^{2}-m_{\rho}^{2}
+m_{\pi}^{2})\over 4\bbox{k}^{2}}
            \right. \right.
\nonumber\\
& & + \left({(E^{2}+\bbox{k}^{2})
\left((M^{2}-m_{\rho}^{2}+m_{\pi}^{2})^{2}
+4\omega_{\pi}^{2}E^{2}\right) \over 32p|\bbox{k}|^{3}} -
{\omega_{\pi}^{2}E^{2}\over 2p|\bbox{k}|} \right. \nonumber\\
& & \left. +{p^{2}(E^{2}+\bbox{k}^{2})+2m_{\pi}^{2}\bbox{k}^{2}\over
8p|\bbox{k}|}
\right)[\ln |a_{\pi}| +i\pi{\cal B}_{\pi}]
\nonumber\\
& & + \left.
{\omega_{\pi}EM^{2}(M^{2}
-m_{\rho}^{2}+m_{\pi}^{2})\over 8p|\bbox{k}|^{3}  }
[\ln |b_{\pi}|-i\pi{\cal B}_{\pi}]   \right]
{1 \over e^{\beta\omega_{\pi}}-1} \nonumber\\
& &\quad\quad\quad\quad\quad\quad\quad\quad + {1 \over \omega_{\rho}}
\left[{-(E^{2}+\bbox{k}^{2})(M^{2}-m_{\pi}^{2}+m_{\rho}^{2})
\over 4\bbox{k}^{2}}
          \right.
\nonumber\\
& & + \left({(E^{2}+\bbox{k}^{2})
\left((M^{2}-m_{\pi}^{2}+m_{\rho}^{2})^{2}
+4\omega_{\rho}^{2}E^{2}\right) \over 32p|\bbox{k}|^{3}} -
{\omega_{\rho}^{2}E^{2}\over 2p|\bbox{k}|} \right. \nonumber\\
& & \left. +{p^{2}(E^{2}+\bbox{k}^{2})+2m_{\rho}^{2}\bbox{k}^{2}\over
8p|\bbox{k}|}
\right)[\ln |a_{\rho}| +i\pi{\cal B}_{\rho}]
\nonumber\\
& & + \left. \left.
{\omega_{\rho}EM^{2}(M^{2}-m_{\pi}^{2}+m_{\rho}^{2})
\over 8p|\bbox{k}|^{3}  }
[\ln |b_{\rho}| -i\pi{\cal B}_{\rho}]  \right]
{1 \over e^{\beta\omega_{\rho}}-1}  \right\rbrace,
\label{eq:gmat}
\end{eqnarray}
and
\begin{eqnarray}
{\cal B}_{\pi} &=& \left\lbrace
{ {1 \quad {\rm if} \ \mu_{\rm \ low} \,
                   \le 2p \le \, \mu_{\rm \ high} \hfill
                    } \atop
  0 \quad {\rm otherwise} \hfill
}
\right.
\end{eqnarray}
\begin{eqnarray}
{\cal B}_{\rho} &=& \left\lbrace
{ {1 \quad {\rm if} \ \nu_{\rm \ low} \,
                   \le 2p \le \, \nu_{\rm \ high} \hfill
                    } \atop
  0 \quad {\rm otherwise} \hfill
}
\right.
\end{eqnarray}
where
\begin{eqnarray}
\mu_{\rm \ low}&=&  \left\vert
  E\sqrt{\left(1+{m_{\pi}^{2}-m_{\rho}^{2}\over M^{2}}
\right)^{2}-{4m_{\pi}^{2}\over M^{2}}}-|\bbox{k}|
\left(1+{m_{\pi}^{2}-m_{\rho}^{2}\over M^{2}}\right)\right\vert \nonumber\\
\mu_{\rm \ high} &=&
    \left\vert E\sqrt{\left(1+{m_{\pi}^{2}-m_{\rho}^{2}\over M^{2}}
\right)^{2}-{4m_{\pi}^{2}\over M^{2}}}+|\bbox{k}|
\left(1+{m_{\pi}^{2}-m_{\rho}^{2}\over M^{2}}\right)\right\vert
\end{eqnarray}
and
\begin{eqnarray}
\nu_{\rm \ low}&=&  \left\vert
  E\sqrt{\left(1+{m_{\rho}^{2}-m_{\pi}^{2}\over M^{2}}
\right)^{2}-{4m_{\rho}^{2}\over M^{2}}}-|\bbox{k}|
\left(1+{m_{\rho}^{2}-m_{\pi}^{2}\over M^{2}}\right)\right\vert \nonumber\\
\nu_{\rm \ high} &=&
    \left\vert E\sqrt{\left(1+{m_{\rho}^{2}-m_{\pi}^{2}\over M^{2}}
\right)^{2}-{4m_{\rho}^{2}\over M^{2}}}+|\bbox{k}|
\left(1+{m_{\rho}^{2}-m_{\pi}^{2}\over M^{2}}\right)\right\vert .
\end{eqnarray}
It is difficult to do little else with these expressions except numerical
analyses.

\subsection{Effective widths, masses and dispersion relations}
\label{sec:widths_and_masses}

Decays are of course observable.  In the Particle Data Group tables
one has experimental values for free-space partial and total
decay widths.  They provide calibration
for our calculation since the partial decay width can be related
to the imaginary part of $F$ (or $G$, since they are the same in vacuum)
as follows
\begin{equation}
\Gamma_{\phi\to \pi\rho} = -{1\over m_{\phi}}{\rm Im}\,F_{\rm vac}
(k^{2}=m_{\phi}^{2}).
\label{eq:phiwidth}
\end{equation}
This uniquely determines the value of $g_{\phi\rho\pi}$.   It is
indeed quite reassuring that the decay width calculated this way is precisely
what one gets from an $S$-matrix approach for the net decay into a
$\pi\rho$ combination, namely,
\begin{eqnarray}
\Gamma_{\phi\to\rho\pi} &=& {1 \over 16}\left({m_{\phi}^{2}
g_{\phi\rho\pi}^{2} \over 4\pi} \right)  {1\over m_{\phi}^{5}}
\left[ \left(m_{\rho}^{2}-m_{\pi}^{2}-m_{\phi}^{2}\right)^{2}
-4m_{\pi}^{2}m_{\phi}^{2} \right]^{3/2}.
\label{eq:rhopiwidth}
\end{eqnarray}
Note that as we are using it, $g_{\phi\rho\pi}$ has units of inverse
mass.  The relevant formula for the two-kaon decay is
\begin{eqnarray}
\Gamma_{\phi\to KK} &=& {2 \over 3}\left({g_{\phi KK}^{2} \over 4 \pi}
\right) {1\over m_{\phi}^{2}}
\left[ {m_{\phi}^{2} \over 4}-m_{K}^{2}\right]^{3/2}.
\label{eq:kakawidth}
\end{eqnarray}
Throughout this work we have used $m_{\pi}$ = 139.6 MeV,
$m_{K}$ = 493.6 MeV, $m_{\rho}$ = 768.1 MeV, $m_{\phi}$ = 1019.4 MeV,
the full width $\Gamma_{\phi}$= 4.43 MeV and partial widths
$\Gamma_{\phi\to KK}/\Gamma_{\phi}$ =
49.1\% and $\Gamma_{\phi\to \rho\pi}/\Gamma_{\phi}$ = 12.9\%.
Equation~(\ref{eq:rhopiwidth}) determines
$m_{\phi}^{2}g_{\phi\rho\pi}^{2}/4\pi$ = 0.19,
while Eq.~(\ref{eq:kakawidth}) determines
$g_{\phi KK}^{2}/4\pi$ = 1.65.

In Figs.~\ref{fig:fig2} and \ref{fig:fig3} we show $-F_{\rm I}$ and
using $k$ = 0.05 GeV/c and $k$ = 0.75 GeV/c. In each
plot, three curves are shown.  Solid, dashed and dotted curves correspond
to temperatures $T$ = 0, 100 and 200 MeV, respectively.  Below
the two-kaon threshold the $\pi\rho$ contribution stands alone and
is noticeably affected by temperature.  The imaginary
part of $G$ behaves almost identically to $F$, so we do not show it.
Equation~(\ref{eq:phiwidth}) also determines an effective width for nonzero
temperatures with the replacement
\begin{equation}
F_{\rm vac} \rightarrow F=F_{\rm vac}+F_{\rm mat}
\end{equation}
This effective width is now dependent not only on the invariant
mass of the resonance, but separately on its three-momentum.  It is
also polarization dependent since instead of $F$ we might have
used $G$.  In practice, this difference is not noticeable.
So using a small value of momentum
we plot the effective widths in
Fig.~\ref{fig:fig4}.  The $\pi\rho$ channel nearly double its
(zero temperature) value at $T$ = 200 MeV.  Since the two-kaon
channel is larger and is affected less by temperature, the sum of
the two channels changes only modestly.  The value for the width
of the sum of both channels goes from 2.74, to 2.90, to 3.22 and finally,
to 3.67 MeV at temperatures 0, 100, 150 and 200 MeV, respectively.

An effective mass can be determined from the pole of
the propagator.  It shifts somewhat due to the presence of matter
since $F$ and $G$ have nonzero real parts.  These parts are in
general different thereby introducing a {\em polarization dependence}.
But in the limit $\bbox{k}\rightarrow 0$, $F$ and $G$ become equal and
one can speak of an effective mass.  Formally, it is the value of $k_{0}$
which satisfies
\begin{equation}
k_{0}^{2} - m_{\phi}^{2} - F_{\rm R}(k_{0},|\bbox{k}|\rightarrow 0,T) =0.
\end{equation}
We extract it from our numerical results.  In Fig.~\ref{fig:fig5}
we show this effective mass minus the vacuum mass as a function
of temperature:  It rises with rising temperature.
At $T=200$ MeV, the effective mass calculated this way increases
by $\sim 4$ MeV.  For finite three-momentum the situation is more
complicated so we show the $K^{+}K^{-}$ and the $\pi\rho$ contributions
to the real part of the function $F$ at two different momenta in
Figs.~\ref{fig:fig6} and \ref{fig:fig7}.  The $\pi\rho$ results are
distinquishable from the two-kaon results since they
extend below the two-kaon
threshold.  For each, three curves are shown:  $T = 0$, 150 and 200
MeV temperature correspond to solid, dashed and dotted lines,
respectively.  The noteworthy features are these.  The $\pi\rho$
contribution is increased at finite temperature for masses ranging
from threshold to roughly 1.3--1.6 GeV (depending on the temperature
and momentum) where it becomes reduced with finite temperature.
Temperature effects are quite small
here.  Kaon contributions are strictly increased at finite
temperature for all masses.  These increases and reductions have
competing effects on the $\phi$ effective mass.

Since $F$ and $G$ are (very) slightly different, a notion more general than
effective mass is needed.  So next we consider the
relationship between energy and momentum for
each polarization, i.e. dispersion.  Evaluation of the dispersions
requires us to find solutions to
\begin{eqnarray}
k_{0}^{2} &=& \bbox{k}^{2} + m_{\phi}^{2} + F_{\rm R \,}(k_{0},|\bbox{k}|,T),
\nonumber\\
k_{0}^{2} &=& \bbox{k}^{2} + m_{\phi}^{2} + G_{\rm R \,}(k_{0},|\bbox{k}|,T),
\end{eqnarray}
for given values of $|\bbox{k}|$ and $T$.  They determine the
longitudinal and transverse dispersion relations $\omega^{\rm L}_{k}$ and
$\omega^{\rm T}_{k}$.  We present them in Figs.~\ref{fig:fig8} and
\ref{fig:fig9} wherein the solid curves are for $T$ = 0, while dashed and
dotted curves correspond to $T$ = 150 and 200 MeV results, respectively.  The
effect is small but the direction
of it is quite clear.  We see an increase in the effective mass with
temperature.  The transverse polarization effects in Fig.~\ref{fig:fig9}
are just slightly greater.

\subsection{thermal dilepton emission rates}
\label{sec:dileptons}

The electron-positron production rate has been shown to
be related to the imaginary part of the retarded photon self-energy
\cite{cgal91}.  Since vector-meson dominance suggests that a virtual photon
converts
first to a neutral vector meson which then couples to other hadrons, the
photon self-energy can be simply related to the imaginary part of the vector
meson propagator.  Derivational details of this equivalence and the
formulas can be found in Refs. \cite{efei76}--\cite{hwel90} and
\cite{cgal91}.
We merely state the formula for the production rate for arbitrary
momentum configurations proceeding through the $\phi$-meson
\begin{eqnarray}
E_{+}E_{-}{dR\over d^{3}p_{+}d^{3}p_{-}} &=&
{1\over (2\pi)^{6}} {e^{4}\over \tilde{g}_{\phi}^{2}}
{m_{\phi}^{4} \over M^{4}} \left\lbrace \left[\bbox{q}^{2}-(\bbox{q}\cdot
\bbox{\hat{k}})^{2}\right] {-F_{\rm I} \over \left(M^{2}-m_{\phi}^{2}-F_{\rm R}
\right)^{2}+F_{\rm I}^{2}}
\right. \nonumber\\
& &+ \left.  \left[2M^{2}-\bbox{q}^{2}+(\bbox{q}\cdot
\bbox{\hat{k}})^{2}\right] {-G_{\rm I}\over \left(M^{2}-m_{\phi}^{2}-G_{\rm R}
\right)^{2}+G_{\rm I}^{2}}
\right\rbrace {1 \over e^{\beta E}-1}
\label{eq:rate}
\end{eqnarray}
where for convenience we have used $q^{\mu}=p_{+}^{\mu}-p_{-}^{\mu}$
and momentum conservation forces $k^{\mu}=p_{+}^{\mu}+p_{-}^{\mu}$.
The rate for $e^{+}e^{-}$ production below the two-kaon threshold
is obtained from Eq.~(\ref{eq:rate}) with $\tilde{g}_{\phi}\rightarrow
m_{\phi}g_{\phi\rho\pi}$.  For invariant masses above threshold,
one adjusts
$\tilde{g}_{\phi}$ to achieve continuity in the rate at $M=2m_{K}$.
Here the real and imaginary parts of $F$ and $G$ have
contributions from both kaon-loop and $\pi\rho$ loop diagrams.
Though we do not show them,
the rates for the individual channels are (branching ratio) fractions
of this total rate.  For example,
\begin{equation}
E_{+}E_{-} {d{R}_{\pi\rho\rightarrow \phi \rightarrow e^{+}e^{-}} \over
d^{3}p_{+}d^{3}p_{-}} =  \left( {F_{I}^{\pi\rho} \over F_{I}^{\pi\rho}+
F_{I}^{K^{+}K^{-}}  } \right)
E_{+}E_{-} {d{R}_{\rm total} \over
d^{3}p_{+}d^{3}p_{-}}
\end{equation}
for longitudinal polarizations.  Then the same expression is used
with the function $F$ replaced by $G$ for
transverse polarizations.

As mentioned several times already, dielectron spectra depend on
polarization.  For example,
\begin{eqnarray}
\bbox{q}^{2}-(\bbox{q}\cdot\bbox{\hat k})^{2} &=& \left\lbrace
{ {M^{2} \quad\quad
  {\rm if} \ \bbox{q} \ \bot \ \bbox{k}
                   \hfill
                    } \atop
   0 \quad\quad\quad {\rm if} \ \bbox{q} \ || \ \bbox{k} \hfill
}
\right. \nonumber\\
2M^{2}-\bbox{q}^{2}+(\bbox{q}\cdot\bbox{\hat k})^{2} &=& \left\lbrace
{ {M^{2} \quad\quad
  {\rm if} \ \bbox{q} \ \bot \ \bbox{k}
                   \hfill
                    } \atop
   2M^{2} \quad\ \, {\rm if} \ \bbox{q}\ || \ \bbox{k} \hfill
}
\right. .
\end{eqnarray}
While at first this seems like a potentially useful tool, we find it is a
tiny effect for these processes.
Shown in Figs.~\ref{fig:fig10} and \ref{fig:fig11}
are the (sum of the two) rates of dielectron production
at two different total momenta $k = $ 0.05 and 0.75 GeV/c.  The solid curve
uses a temperature independent form factor and the longitudinal and
transverse contributions are presented as dashed and dotted lines,
respectively.  Keeping with the above statement about polarizations, we
note that though they are truly distinct especially at higher momentum,
one probably cannot expect to utilize this difference in experiment.
Another way to think of the effective
mass is that it is the average invariant mass of all back-to-back
kinematically allowed configurations for the dileptons.
If one imagines doing such an average in Fig.~\ref{fig:fig10}, one can see
the rise.  Both the kaons' and the $\pi\rho$\,'s contributions
tend to raise the effective mass.

\section{Conclusions}
\label{sec:conclusions}

It is not possible to overstate the importance of understanding
hadronic processes whitin hot matter.  As the temperature
rises above say, 150 MeV and approaches 200 MeV, hadronic interactions
may well deviate drastically from vacuum behavior.
These modifications are quite important both
from theoretical and experimental points of view.
Open questions abound with regard to the approach to chiral symmetry
restoration and deconefinement phase transition.  A
solid understanding of hadronic physics is essential for making progress
in this direction.
We have studied how temperature affects properties of the $\phi$-meson.
By evaluating a new contribution to the one-loop self-energy,
we have improved estimates for the decay width, dispersion relations
or effective mass, and finally, for the rate of dielectron production
with invariant masses near the $\phi$.

We have learned that within this type of modelling for the relevant
hadronic degrees of freedom and their (strong) interaction, the
effective mass of the $\phi$-meson is nearly temperature independent below
$\sim$ 50 MeV.  Above this, it shows slight increase:  It rises by
0.42, 1.40, and 4.32 MeV at temperatures
100, 150 and 200 MeV, respectively.
This rise is mostly due to the kaon contributions, although
the effect of the $\pi\rho$ channel is to move the mass in the same
direction.  These results for the effective mass contrast with
some QCD sum-rule calculations at finite (temperature and)
density \cite{KoAs}, where a large decrease of
the $\phi$-meson mass at $T \ne 0$, e.g. $\sim$ 100 MeV reduction
at $T=175$ MeV.  However, it is vital in such calculations to treat the
spectral-function side of the sum rule consistently with the change
of the condensate appearing in the operator product expansion.
Moreover, the absence of the $O(T^{2})$ mass shift is quite a general
result on the basis of chiral symmetry~\cite{{VLE93},{yujiprd}}.
Koike~\cite{yuji93} further points out in the context
of QCD sum rules for the vector mesons at finite density (in
nuclear matter) that the medium (density) dependent part of the
correlation function has to be identified from the point of view
of the forward current-nucleon scattering amplitude.  This statement
also applies to the finite-$T$ sum rule for the $\phi$-meson~\cite{yuji94}. In
support of approaches like ours we wish to stress here its virtues of
self--consistency: the thermal modifications of the real and of the imaginary
parts of the $\phi$ self--energy are computed simultaneously, not just one or
the other. This has considerable appeal for the point of view of theory.

The partial decay width into
the $\pi\rho$ channel is the most affected by temperature.  Since it
is smaller than the kaon channel to begin with, the net result for
the width's modification is also modest.  At 150 MeV temperature,
the width increases by 25\% and at 200 MeV temperature it increases
by 34\% as compared to its vacuum value.
The thermal rate of dielectron production near the $\phi$ mass
was shown to be altered accordingly.  Polarization
effects are insignificantly small.

While our study answers some questions about temperature's affect
on properties of the $\phi$-meson, it does not address some potentially
important issues.  In the context of a loop expansion of the photon
self--energy, collisional broadening effects on the dilepton spectrum are of
two--loop order, beyond the present calculation. However, those have recently
been estimated \cite{Kosei}. A consistent incorporation of such contributions
in our self--consistent framework is desirable and considered. Probably the
next-most-important idea to incorporate
into this calculation is a modified pion dispersion relation.
Pion properties are probably also modified at such high temperatures
and it is important to see how sensitive other hadrons'
properties are to such things.  There is also the question of
finite chemical potentials.  We have neglected it,
but inclusion of a nonzero $\mu_{\pi}$ is rather straightforward.
These activities are planned.

\section*{Acknowledgements}

Our research is supported in part by the National Science Foundation
under grant number PHY-9017077, by the Natural Sciences and Engineering
Research Council of Canada, by the FCAR of the Qu\'ebec Government and
by a NATO Collaborative Research Grant.

\begin{figure}
\caption{Diagrams that contribute to the one-loop $\phi$
self-energy.  The kaon bubble and tadpole are
shown in (a) and (b) where the dotted circles represent kaons,
while (c) shows the $\pi\rho$ loop.}
\label{fig:diagrams}
\end{figure}
\begin{figure}
\caption{The imaginary part of the function $F$ at
three-momentum $k$ = 0.05 GeV/c as it depends on the invariant
mass.  The solid curve is the vacuum ($T$=0) contribution, the dashed
and dotted curves correspond to  $T$ = 100 and 200 MeV, respectively.}
\label{fig:fig2}
\end{figure}
\begin{figure}
\caption{Same as in Fig.\ \protect\ref{fig:fig2}
except a higher three-momentum of $k$ = 0.75 GeV/c.}
\label{fig:fig3}
\end{figure}
\begin{figure}
\caption{Partial decay widths as a function of temperature.}
\label{fig:fig4}
\end{figure}
\begin{figure}
\caption{The effective mass minus the vacuum mass
$m^{\rm eff}_{\phi} - m^{\rm vac}_{\phi}$ versus temperature.}
\label{fig:fig5}
\end{figure}
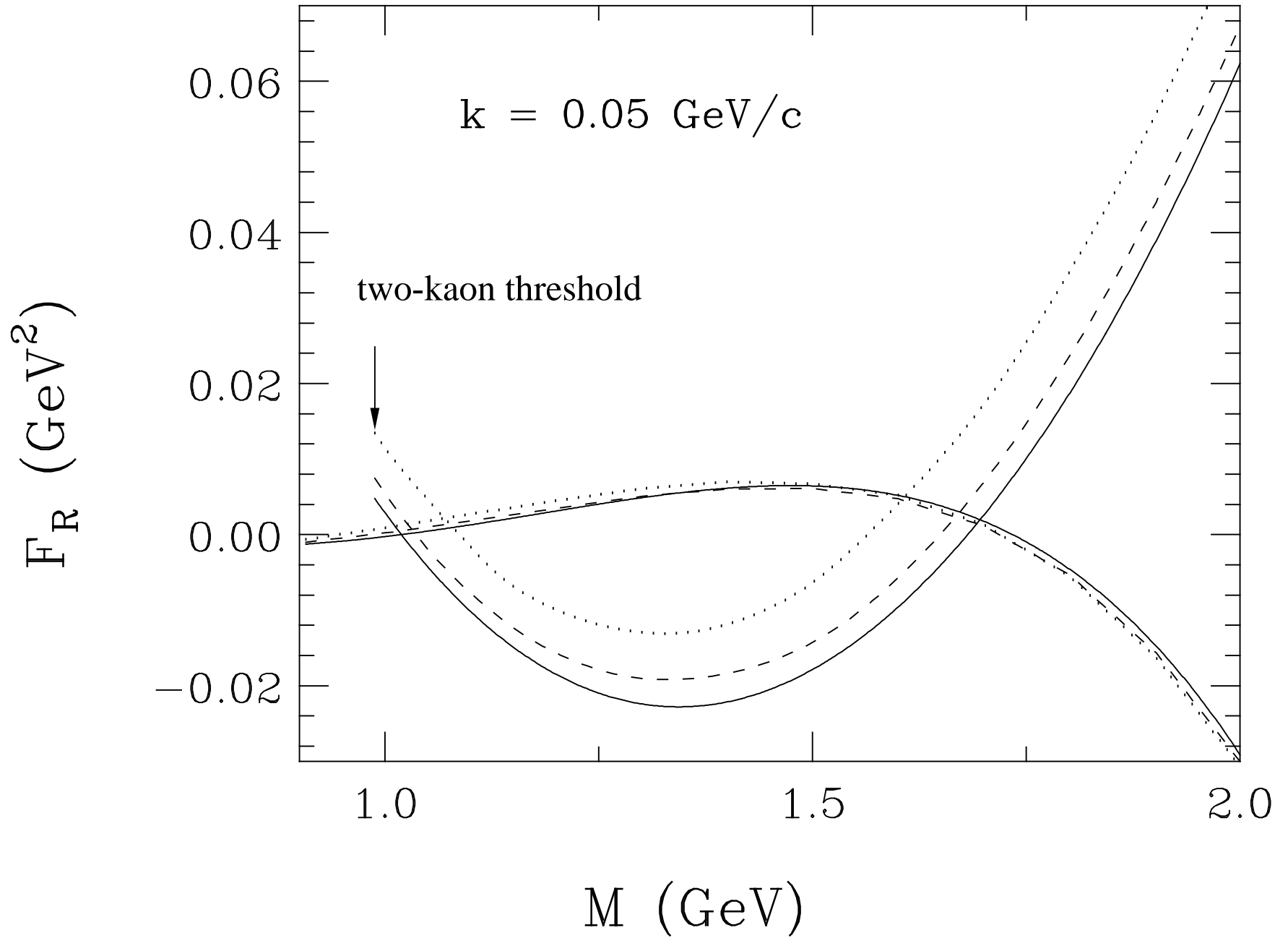
\begin{figure}
\caption{The real part of $F$ at fixed
three-momentum $k$ = 0.05 GeV/c as a function of the invariant mass.
Kaon contributions correspond to the three curves that begin at the
$2m_{K}$ threshold as indicated.  The $\pi\rho$ contributions are
the three other curves on which temperature has less affect.
 }
\label{fig:fig6}
\end{figure}
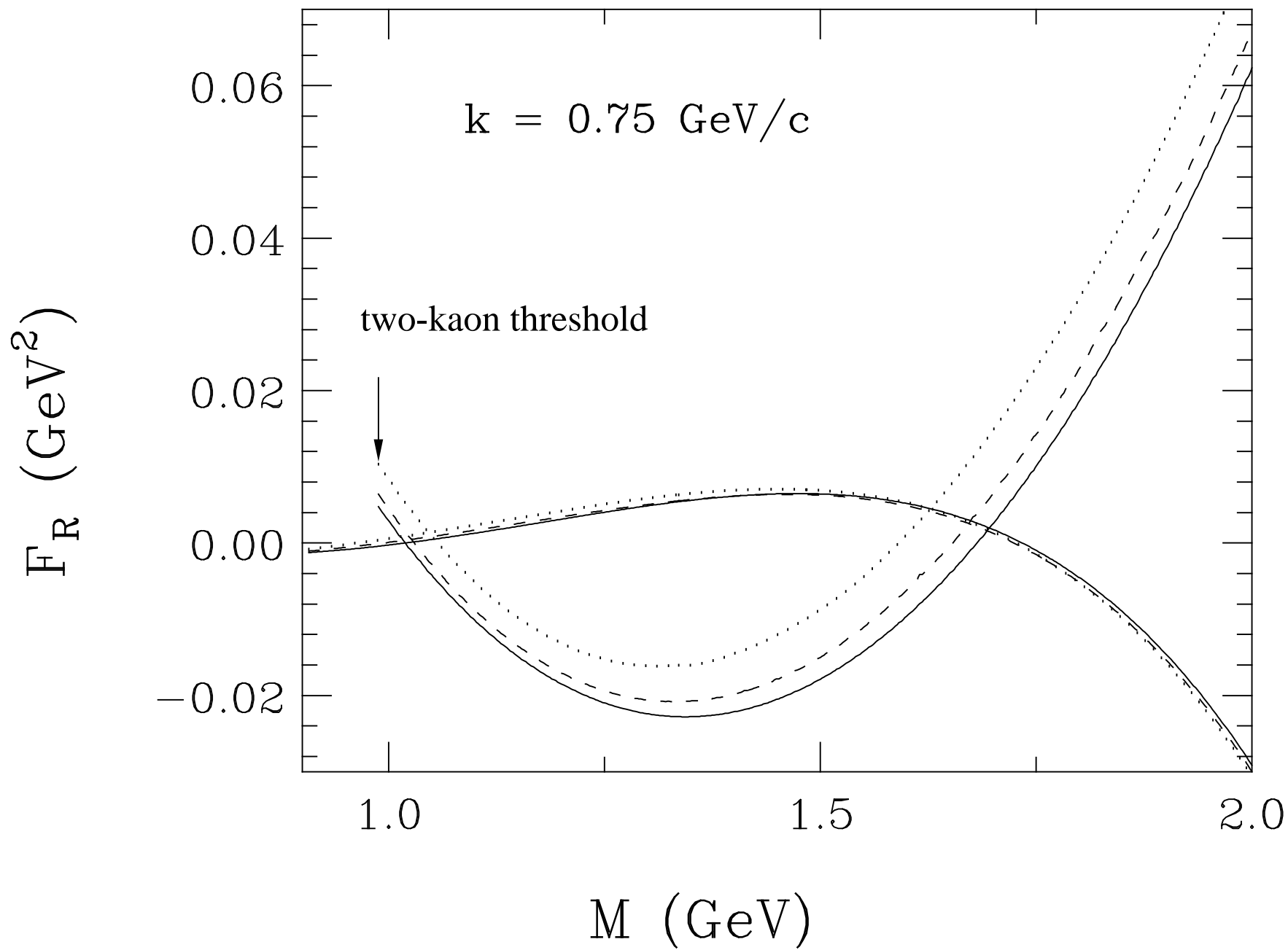
\begin{figure}
\caption{The same as Fig.\ \protect\ref{fig:fig6}
with momentum $k$ = 0.75 GeV/c.}
\label{fig:fig7}
\end{figure}
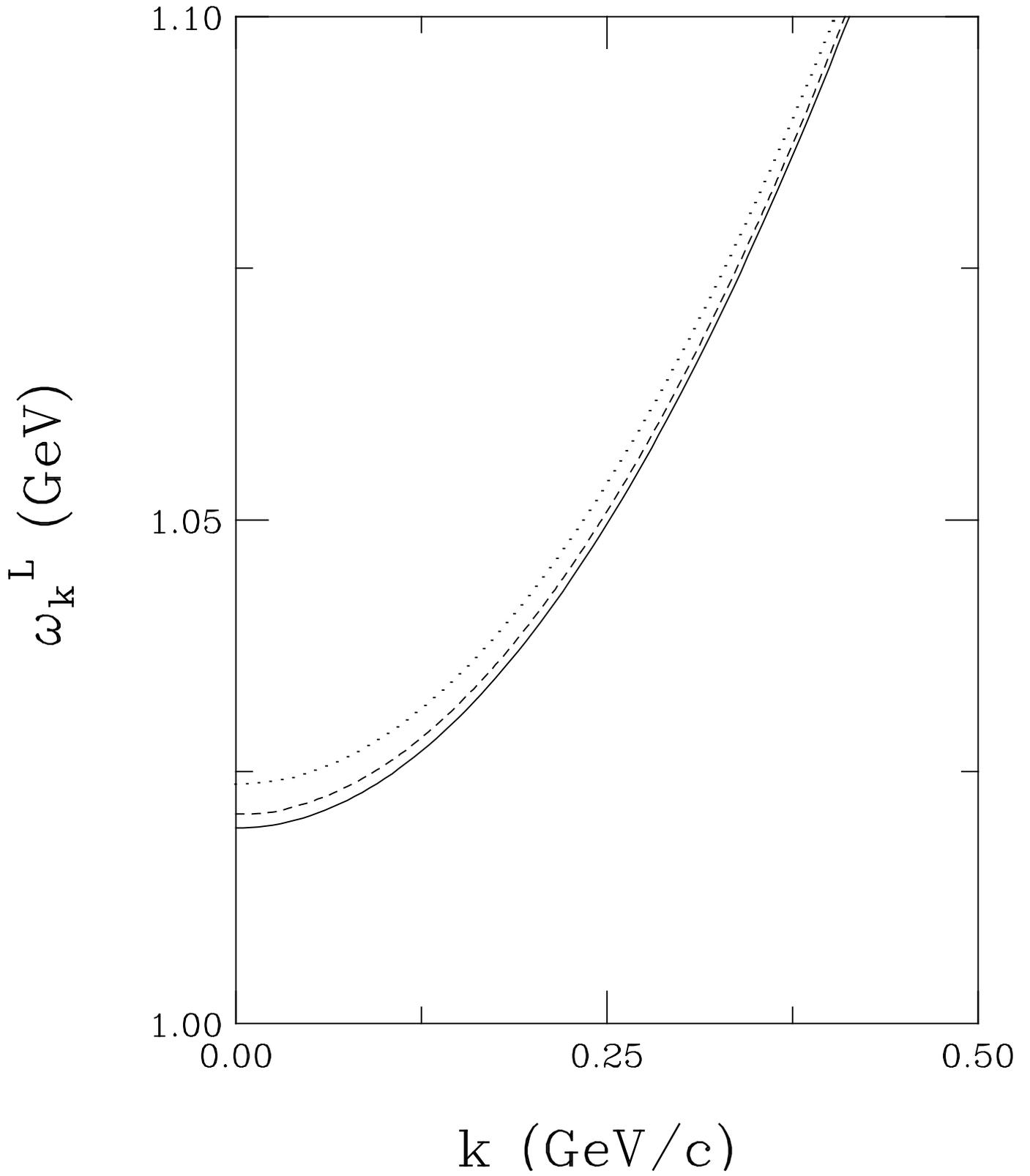
\begin{figure}
\caption{Dispersion relation for longitudinally polarized
$\protect\phi$-mesons at zero (solid curve) and finite temperature (dashed
curve $T$=150 MeV and dotted curve $T$= 200 MeV).}
\label{fig:fig8}
\end{figure}
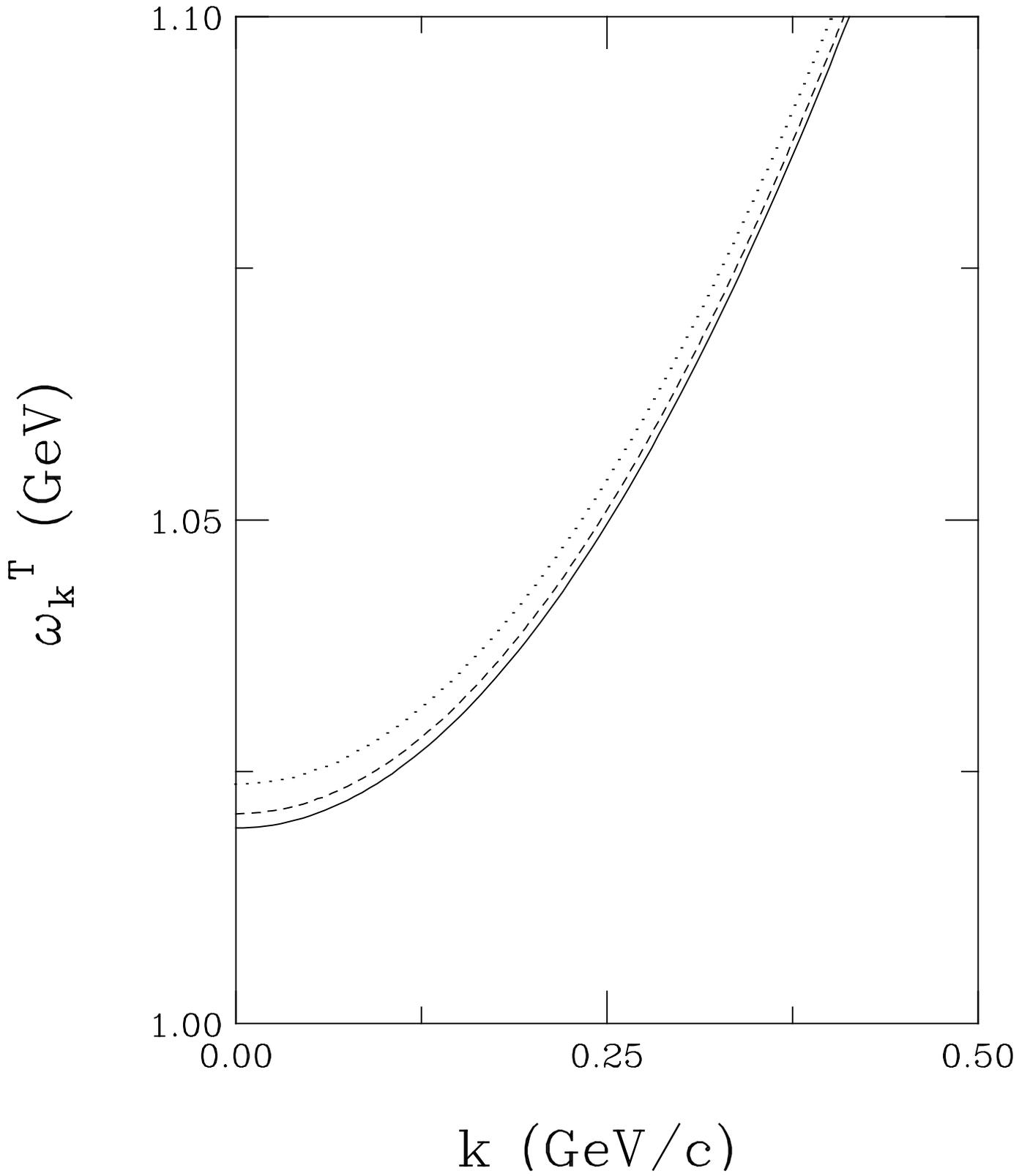
\begin{figure}
\caption{Same as Fig.\ \protect\ref{fig:fig8}
except for transversely polarized $\protect\phi$-mesons.}
\label{fig:fig9}
\end{figure}
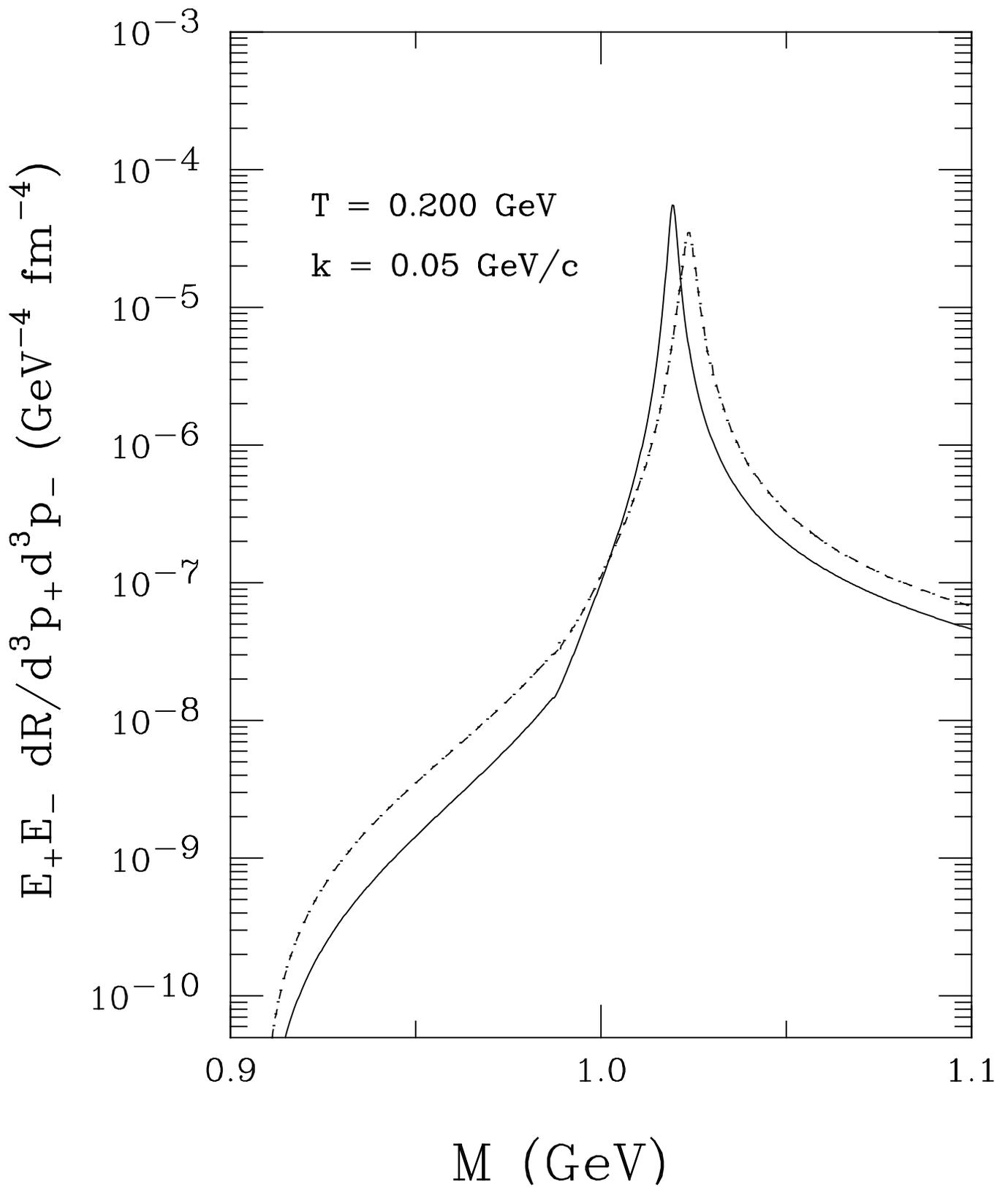
\begin{figure}
\caption{Dielectron production rate for invariant masses near $m_{\phi}$
from the sum of $\protect K^{+}K^{-} \rightarrow \phi\rightarrow e^{+}e^{-}$
and $\protect \rho\pi \rightarrow \phi\rightarrow e^{+}e^{-}$.
The solid curve uses a $T$-independent form factor, whereas the dashed curve
is for longitudinal polarizations and the dotted curve is for transverse
polarizations.  The value of three-momentum and temperature
are $k$ = 0.05 GeV/c and 200 MeV.}
\label{fig:fig10}
\end{figure}
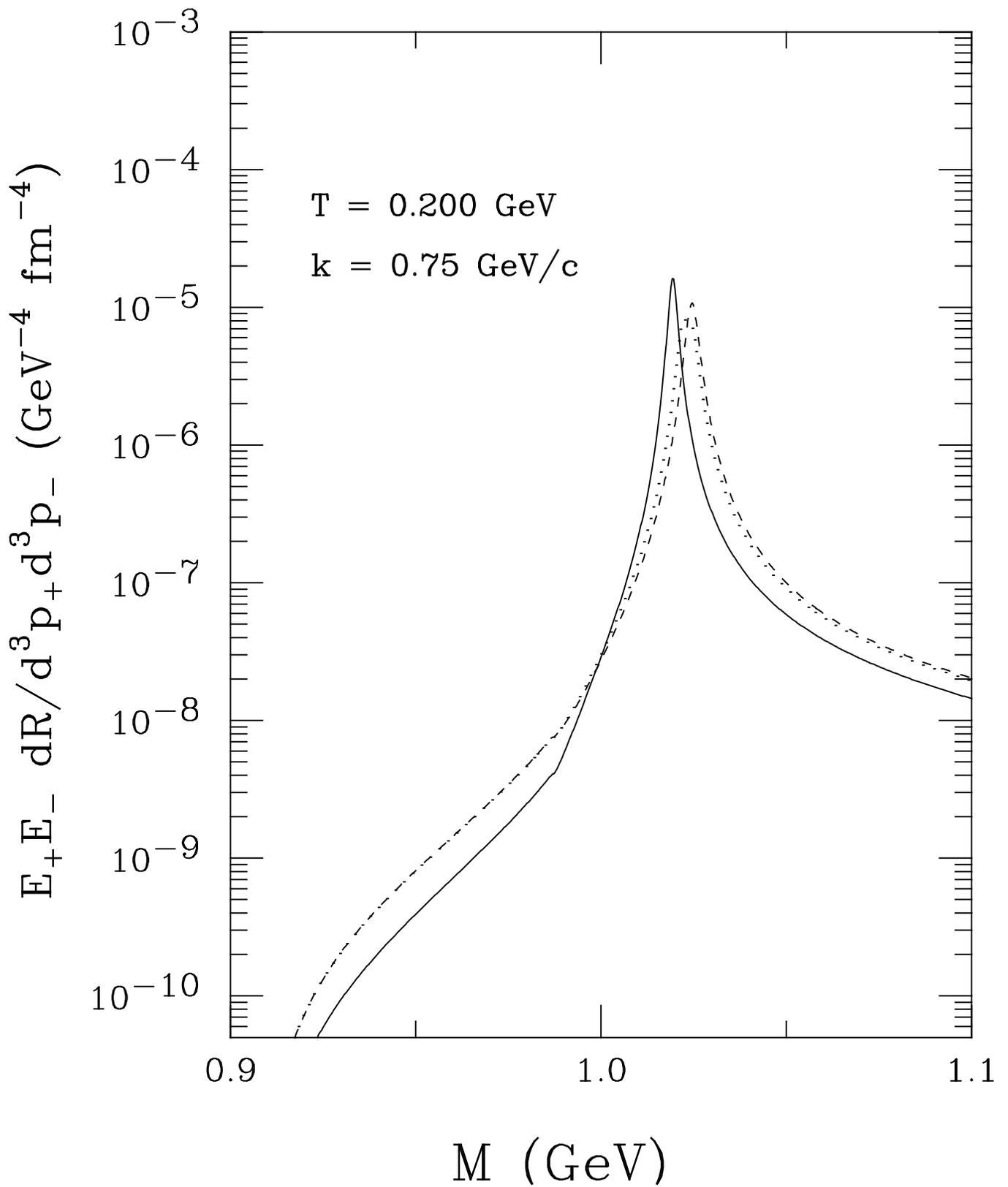
\begin{figure}
\caption{Same as Fig.\ \protect\ref{fig:fig10} except momentum
$k$ = 0.75 GeV/c.}
\label{fig:fig11}
\end{figure}

\end{document}